\documentclass[conference,usenatbib]{basi}
\usepackage[T1]{fontenc}
\usepackage[british]{babel}
\usepackage[varg]{txfonts}
\usepackage{rotating}
\usepackage{dcolumn}
\begin{document}
\newcommand{\apj}{ApJ}
\newcommand{\apjs}{ApJS}
\newcommand{\apjl}{ApJL}
\newcommand{\aap}{A{\&}A}
\newcommand{\aaps}{A{\&}AS}
\newcommand{\mnras}{MNRAS}
\newcommand{\aj}{AJ}
\newcommand{\araa}{ARAA}
\newcommand{\pasp}{PASP}
\newcommand{\pasj}{PASJ}
\newcommand{\nar}{New Astr. Rev.}
\newcommand{\ssr}{Space Sci. Rev.}
\newcommand{\cjaa}{CJAA}
\newcommand{\na}{New Astr. Rev.}
\newcommand{\nat}{Nature}
\newcommand{\apss}{AP{\&}SS}
\newcommand{\fermi}{{\it Fermi}}
\newcommand{\nustar}{{\it NuSTAR}}
\title[Lepto$-$Hadronic Modeling of 3C 279]{Lepto$-$Hadronic Origin of $\gamma$-ray Outbursts of 3C 279}
\author[Paliya et~al.]%
       {Vaidehi S. Paliya$^{1,2}$\thanks{email: \texttt{vaidehi@iiap.res.in}},
       Chris Diltz$^{3}$, Markus B{\"o}ttcher$^{4,3}$, C. S. Stalin$^{1}$, and David Buckley$^{5}$\\
       $^1$Indian Institute of Astrophysics, Block II, Koramangala, Bangalore-560034, India\\
       $^2$Department of Physics, University of Calicut, Malappuram-673635, India\\
       $^3$Astrophysical Institute, Department of Physics and Astronomy, Ohio University, Athens, OH 45701, USA\\
       $^4$Centre for Space Research, North-West University, Potchefstroom, 2520, South Africa\\
       $^5$South African Astronomical Observatory, PO Box 9, Observatory 7935, Cape Town, South Africa}

\pubyear{2015}
\volume{12}
\pagerange{\pageref{firstpage}--\pageref{lastpage}}

\date{Received --- ; accepted ---}

\maketitle
\label{firstpage}

\begin{abstract}
The blazar 3C 279 exhibited a giant $\gamma$-ray outburst in 2013 December and 2014 April. Apart from the very fast $\gamma$-ray flux variability, the spectral nature of the flares were also found to vary significantly between these two flaring events. A prominent curvature in the $\gamma$-ray spectrum was noticed in 2014 April flare, on the other hand, the 2013 December displayed an extreme hardening of the spectrum. These observations, thus, put strong constraints on our understanding of the underlying particle acceleration mechanisms.

\end{abstract}


\section{Introduction}\label{s:intro}
The quasar 3C 279 ($z$ = 0.536) was detected in a high state in 2013 December by \fermi~satellite. During this period, not only an extremely bright $\gamma$-ray flare was observed, but also the detected $\gamma$-ray spectrum was hard. In this work, this peculiar flare is studied.
\section{Results and Conclusions}
\begin{figure}
\centerline{
\includegraphics[width=7cm]{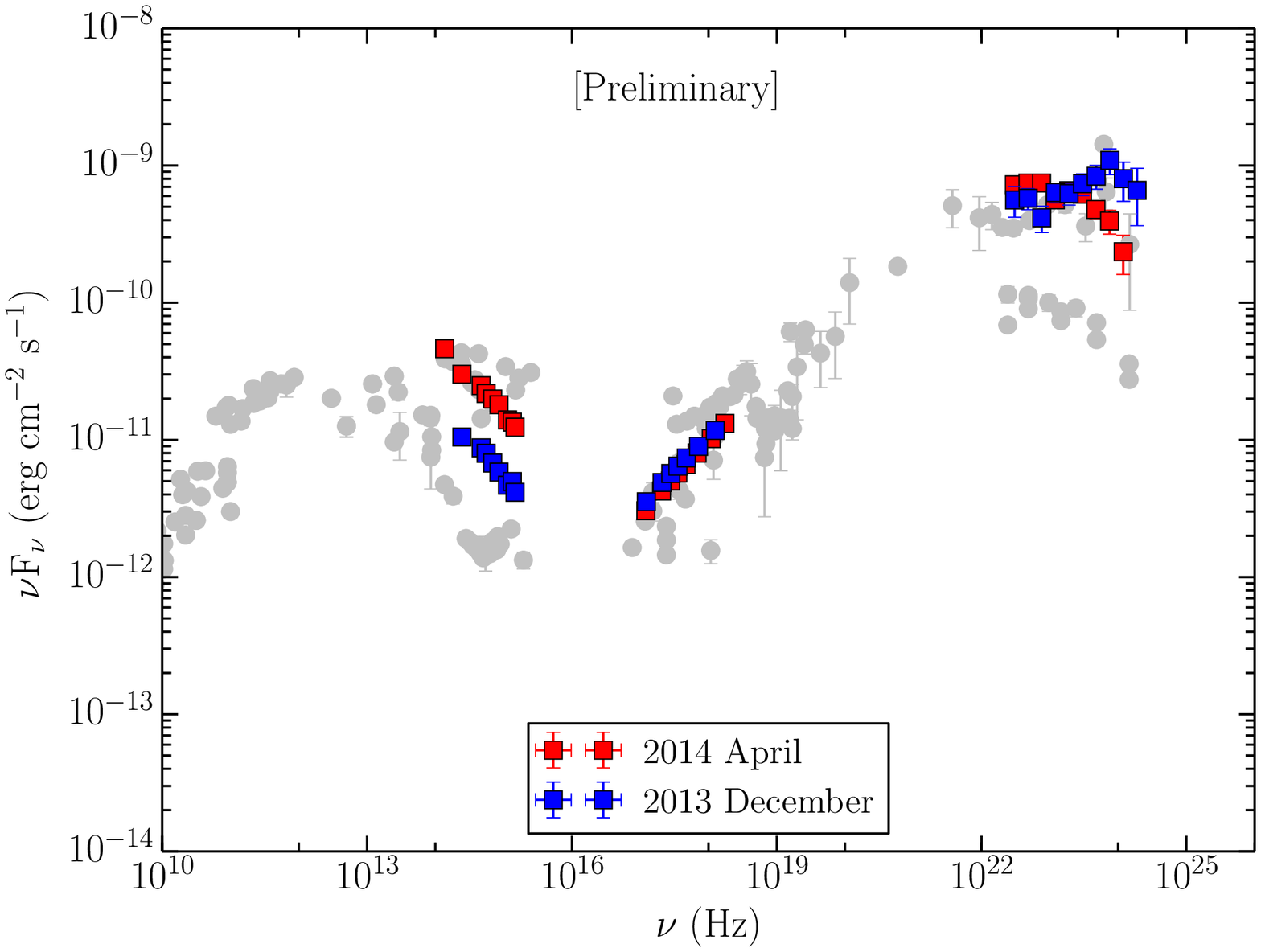}
\includegraphics[width=7cm]{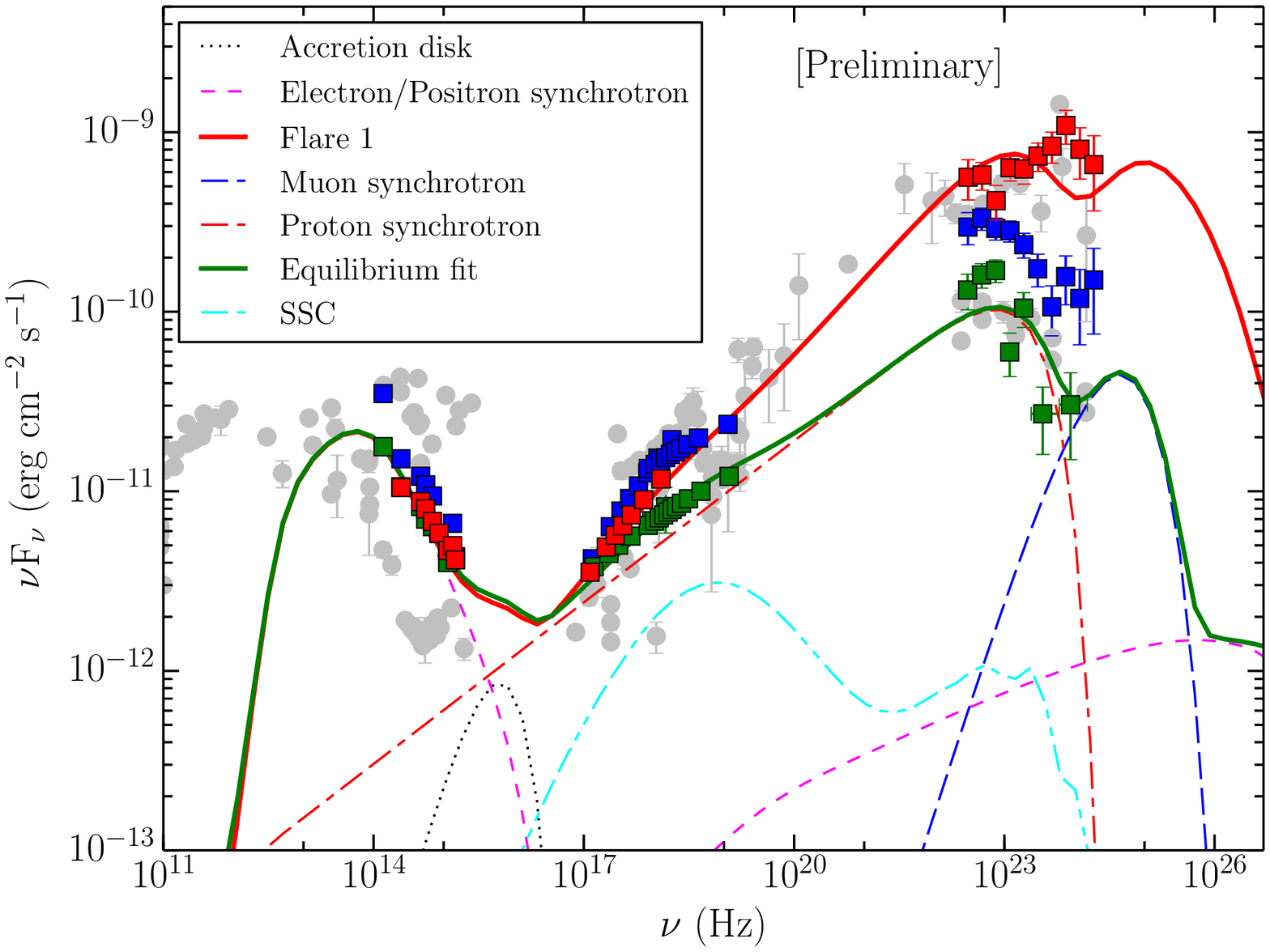}
}
\caption{Left: Comparison of the SEDs of 3C 279 at the peak of the flares in 2013 December (F1) and 2014 April (F2). Right: Time dependent lepto-hadronic model fits to the SED of 3C 279.\label{fig:SED_Apr_Dec}}
\end{figure}

In the left panel of Figure~\ref{fig:SED_Apr_Dec}, we compare the SEDs of 3C 279 covering the 
2013 December and 2014 April flares (hereafter F1 and F2, respectively). As can be seen,
 the shape of the optical and X-ray SEDs remains same during both the flares, but what is more appealing is the change in the shape of $\gamma$-ray spectrum. In one zone leptonic scenario, the shape of the synchrotron spectrum constrains the shape of the high energy $\gamma$-ray radiation, assuming the same population of electrons are emitting both emissions \citep[see e.g.,][]{2012MNRAS.419.1660S}. Accordingly, a falling optical spectrum should correspond to the steep $\gamma$-ray spectrum, which is observed in F2. However, a rising $\gamma$-ray spectrum, as seen during F1, is difficult to explain with this interpretation, since optical spectrum is declining. This observation indicates that a single zone leptonic model cannot explain the hard $\gamma$-ray flare detected in F1. Therefore, we reproduce these peculiar observations following a time dependent lepto-hadronic emission scenario described in \citet{2015ApJ...802..133D}. The result of this modeling is shown in the right panel of Figure~\ref{fig:SED_Apr_Dec}. The model first reaches the equilibrium (solid green line) and then injects a Gaussian perturbation to the input parameters to reproduce the flare (solid red line). The $\gamma$-ray data of the high state is not fitted very well, but keeping in mind that the input perturbations try to reproduce the overall SED (including IR to X-ray spectra), we believe that the model is a good representation of the data.

The multi-wavelength observations of 3C 279 during the twin $\gamma$-ray flares have revealed that F1 cannot be explained by one-zone radiative models and we need to look for alternative approach such as lepto-hadronic modeling. These observations hint for the presence of a variety of radiative processes working at different activity levels in the jet of 3C 279, of which, these two flares are just an example.

\end{document}